\documentclass[reprint, showpacs,aps, prl]{revtex4-1}
\usepackage{dcolumn}
\usepackage[english]{babel}
\usepackage{amsmath,graphicx,xcolor,hyperref,picture,calc}
\usepackage{epstopdf}

\catcode`\<=\active \def<{
\fontencoding{T1}\selectfont\symbol{60}\fontencoding{\encodingdefault}}
\catcode`\|=\active \def|{
\fontencoding{T1}\selectfont\symbol{124}\fontencoding{\encodingdefault}}

\begin{document}

\title{Reordering Fractional Chern Insulators into Stripes of Fractional Charges with Long-Range Interactions}

\author{Mengsu Chen}

\author{V. W. Scarola}
\affiliation{Department of Physics, Virginia Tech, Blacksburg, Virginia 24061
USA}

\date{\today}

\begin{abstract}
Long-range interactions drive some of the rich phenomenology of quasiparticle collective states in the fractional quantum Hall (FQH) regime.  We test for analogues in models of fractional Chern insulators (FCIs) derived from a screened Coulomb interaction.  We find that the uniform FCI liquid is surprisingly robust to long-range interactions but gives way to a unidirectional charge density wave (CDW) of fractionally charged quasiparticles with increased screening length.  Our results show that FCIs offer a robust and important platform for studying quasiparticles collective states.
 \end{abstract}
 
 \pacs{73.21.-b, 03.65.Vf, 71.27.+a, 73.43.-f}
 
{\maketitle}

The collective behavior of quasiparticles can lead to non-Fermi liquids that set new paradigms.  The FQH regime offers one of the best known examples \cite{tsui:1982, laughlin:1983}.  Here uniform quantum liquids of fractionally charged quasiparticles, called composite fermions (CFs) \cite{jain:1989}, form at certain fractional Landau level (LL) fillings.  Wavefunction analyses of CF exciton \cite{scarola:2000} and Cooper pair instabilities \cite{scarola:2000a} show that FQH liquids transition to other intriguing quasiparticle states: crystals \cite{yi:1998} and paired states \cite{moore:1991}, respectively.  CFs have even been argued to form, much like electrons \cite{fogler:2002,fogler:1996,cote:2000}, their own stripe states \cite{lee:2001,lee:2002}. These studies show that some of the richness of the quasiparticle phase diagram in the FQH regime derives from the long-range part of the Coulomb interaction between electrons. 

Recent work on short-range Hubbard models found an intriguing analogue of FQH states in a flat band but in the absence of a net magnetic field \cite{neupert:2011, sheng:2011,bernevig:2011,wu:2012a,lauchli:2013,liu:2013a,parameswaran:2013,bergholtz:2013}.  This uniform quantum liquid, the FCI, should also display its own FQH effect derived from fractionally charged quasiparticles.  Refs.~\cite{grushin:2012,kourtis:2012b} have further shown that increasing the single-particle bandwidth drives the uniform FCI state into a CDW of the original particles via nesting.  

Since FCI and FQH models are adiabatically connected \cite{hai:2012}, we can understand such transitions by appealing to the two distinct CDWs considered in the FQH regime: CDWs of electrons and CDWs of CFs.  CDWs of electrons at filling $\nu$ of higher LLs are well approximated by Hartree-Fock analyses in the electron degrees of freedom \cite{fogler:2002,fogler:1996,rezayi:1999,cote:2000}, with, e.g., a Hartree-Fock wavefunction: $\phi_{\nu}$, dictated by the electron-electron interaction.  The FCI-CDW transition found in Refs.~\cite{grushin:2012,kourtis:2012b} is similar to transitions between FQH liquids and CDWs of electrons in higher FQH LLs \cite{fogler:2002,fogler:1996,cote:2000}.  But in lower LLs, CFs can form distinct CDWs \cite{yi:1998,lee:2001,lee:2002,jain:2007} captured by wavefunctions \cite{jain:1989,yi:1998,lee:2001,lee:2002}: $J^{2p}F_{\rho_{\bf Q}}\phi_{\nu^{*}}^{\text{CF}}$, that are not perturbatively connected to electron CDW wavefunctions.  Here the Jastrow factor $J^{2p}$ attaches $2p$ vortices to each electron to yield a wavefunction at filling $\nu^{*}/(2p\nu^{*}+1)$.  $\phi_{\nu^{*}}^{\text{CF}}$ is a uniform Hartree-Fock wavefunction of CFs.  Inter-CF interactions can cause CDW ordering that favors application of the density wave operator, $\rho_{\bf Q}$, in certain combinations, defined by the function $F_{\rho_{\bf Q}}$, to generate a CDW of CFs ordered at wavevector ${\bf Q}$ \cite{lee:2001,lee:2002,jain:2007}.  It is currently unknown if FCI quasiparticles can themselves form CDWs to define rich phase diagrams akin to what has been found for CFs in the lowest LL. 

Stability to long-range interactions and screening defines a crucial difference between the FCI and FQH regimes.  In the FQH regime, only the {\it short}-range part of the bare electron-electron interaction is screened, leaving a Coulomb tail \cite{macdonald:1984,zhang:1986}. But if flat bands defining FCIs are to be found in materials, the basis must be defined by band structure effects (e.g., interactions in combination with multi-orbital states \cite{kourtis:2012a,kourtis:2012b}), not strong magnetic fields.  The relevant FCI interaction to study is therefore a Coulomb interaction that has its {\it long}-range part screened.  It is currently unknown if FCIs and related quasiparticle collective states are too unstable, even for screened interactions, to be found in nature.  But promising work \cite{yao:2013,liu:2013,kourtis:2014} suggests that FCIs might be robust.  

In this Letter we study a screened Coulomb interaction in the FCI regime to explore stability of the quantum liquid and search for collective behavior of quasiparticles.  We project a Yukawa potential, with a screening length $\lambda$, into the flat band used to define the FCI \cite{sun:2011,parameswaran:2013}.   We find that the FCI is surprisingly stable for $\lambda$ as large as 6 lattice constants.  We also find a transition to a unidirectional CDW, a stripe phase, as the screening length is increased into the Coulomb limit.  

The stripe phase we find here is distinct from stripe phases of electrons normally discussed in the FQH regime in high LLs.  We find that the instability to the stripe phase in the FCI Coulomb model is driven by the rearrangement of fractionally charged quasiparticles, from a uniform liquid into the stripe phase.  The stripe phase we find here is akin to the stripe phases of CFs studied in the lowest LL driven by inter-CF interactions \cite{lee:2001,lee:2002} and complements recent work on FCIs \cite{kourtis:2014a}.  We show this conclusively by explicitly computing the charge of each stripe to find an exact rational fraction.  Our findings have important consequences in the search for FCI physics in materials because we show that they are stable for large screening lengths and we have found at least one intriguing collective state of quasiparticles, a stripe state, derived from excitations of the FCI state.  

\noindent
{\it Model:}  We consider a single-particle basis derived from the following model on the checkerboard
lattice \cite{sun:2011}: $ \sum_{{\bf k},m} \psi^{\dagger}({\bf k})   d_{m} ({\bf k} ) \sigma_m \psi({\bf k})$, 
where the fermion pseudo-spinor basis states are defined on sublattices $A$ and $B$ in the 2-site unit cell: $\psi({\bf k})\equiv(
   c_{{\bf k},A}  , 
   c_{{\bf k},B} )^{T}$, the flat band Hamiltonian parameters are $d_1 ( {\bf k} ) =2 \sqrt{2}
\cos ( k_x /2 ) \cos ( k_y /2 )$, $d_2 ( {\bf k} ) =2 \sqrt{2} \sin
( k_x /2 ) \sin ( k_y /2 )$, $d_3 ( {\bf k} ) =(2-\sqrt{2})[ \cos ( k_x ) - \cos
( k_y ) ]$, and $\sigma$ are the Pauli matrices.  We work in the lowest flat band but we have checked that including a finite bandwidth does not qualitatively change the results presented below.

Recent work found numerical evidence for a gapped uniform quantum liquid at a filling of $1/3$ by projecting the nearest neighbor Hubbard interaction into the lowest flat band defined by $d_m$ \cite{neupert:2011,sheng:2011,bernevig:2011}.  The resulting state, in direct analogy to FQH states, demonstrate a 3-fold topological degeneracy, a fractionalized Chern number and should, in principle, exhibit a FQH effect if found in materials.  The role of long-range interactions derived from the Coulomb interaction remains a key issue.

To study the interplay between screening and FCIs we consider a model defined by a Yukawa interaction:
\begin{equation}
H=\mathcal{P}  \frac{e^{2}}{2\epsilon}  \sum_{i \neq i'}\frac{e^{-r_{i  i'} / \lambda}}{r_{i i'}} n_i n_{i'} \mathcal{P},
\label{eq_H}
\end{equation}
where the projectors $\mathcal{P}$ imply that the model acts only in the Fock space of the lowest single-particle flat band and $n_i$ is the fermion number operator. Both $r_{i i'}$ (the inter-site separation before projection) and $\lambda$ are in units of the spacing between the unit cells.  We set the energy unit to $V\equiv \exp(-1/\sqrt{2}\lambda)[\sqrt{2}e^{2}/\epsilon ]$ to ensure that the $\lambda\rightarrow 0$ limit of $H$ reduces to the nearest neighbor interaction while $\lambda\rightarrow \infty$ reduces to the Coulomb limit.  $\epsilon$ is the dielectric constant.  In the following we study the eigenstates of Eq.~(\ref{eq_H}) at $N/N_{c}=1/3$ filling on a periodic $L_{x}\times L_{y}$ lattice, where $N_{c}$ is the number of unit cells.

\noindent
{\it Energetics:} We numerically diagonalize Eq.~(\ref{eq_H}) using the Lanczos algorithm.  
We study $N=6$, $8$, $10$, and $12$ and present results for $L_{x}\times L_{y}=(N/2)\times6$.  We are limited to $N\leq12$ because larger Hilbert space sizes ($\sim 1\times10^{9}$ states for $N=12$) are prohibitive.  We studied other aspect ratios and have also considered tilted periodic boundary conditions which, only for $N=6$ at this filling, allows us to access an additional system size.  In all systems studied we find uniform FCI states with the requisite 3-fold topological degeneracy and a robust gap, $\Delta_{\lambda}$, in the nearest neighbor limit, $\lambda\rightarrow 0$.  But the Coulomb limit reveals a different state.

\begin{figure}[t]
  \resizebox{1\columnwidth}{!}{\includegraphics{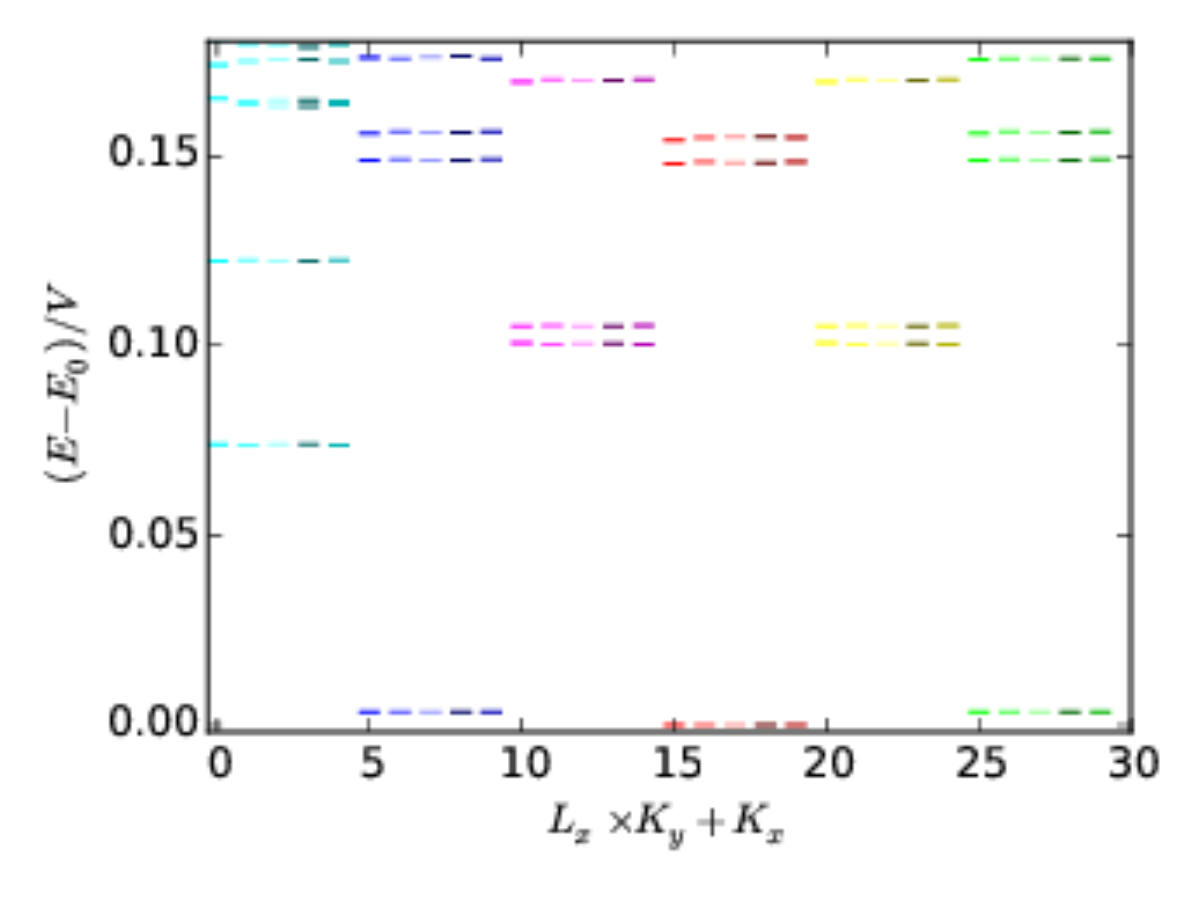}}
  \caption{Energy spectrum of Eq.~(\ref{eq_H}) for $N=10$ particles for the Coulomb
  interaction ($\lambda = \infty$) plotted as a function of integers $(K_{x},K_{y})$, where the total momentum is $2\pi(K_{x}/L_{x},K_{y}/L_{y})$. The ground state energy is $E_{0}$.  For comparison, the FCI state occurs at $L_{x}\times K_{y}+K_{x}=5,15,$ and $25$ for the short range interaction.  The 15-fold degenerate stripe state manifold therefore includes the states at the same wavevectors as the FCI.  }
  \label{E_vs_k}
\end{figure}

Fig.~\ref{E_vs_k} shows representative data for the eigenvalues of Eq.~(\ref{eq_H}) in the Coulomb limit.  Similar results were obtained for all accessible $N$ and large $\lambda$.  
The 15-fold degeneracy found here stems from a 5-fold increase in the 3-fold degeneracy expected for the FCI state.  In general, we find that the degeneracy increases to $3\times L_{x}$ ($3\times 2$) for $L_{x}$ odd (even).  The degeneracies occur at wavevectors corresponding to a stripe modulation of the underlying uniform FCI states: $K_{FCI}+(n_{x},0)$, where $K_{FCI}$ is any of the 3 topologically degenerate wavevectors of the FCI state, $n_{x}=0$ and $L_{x}/2$ when $L_{x}$ is even (i.e., when $L_{x}$ an integer multiple of the stripe spacing), and $n_{x}=0,1,...,L_{x}-1$ otherwise.  We discuss these degeneracies in the following sections.

\begin{figure}[t]

  \includegraphics[width=3.5in]{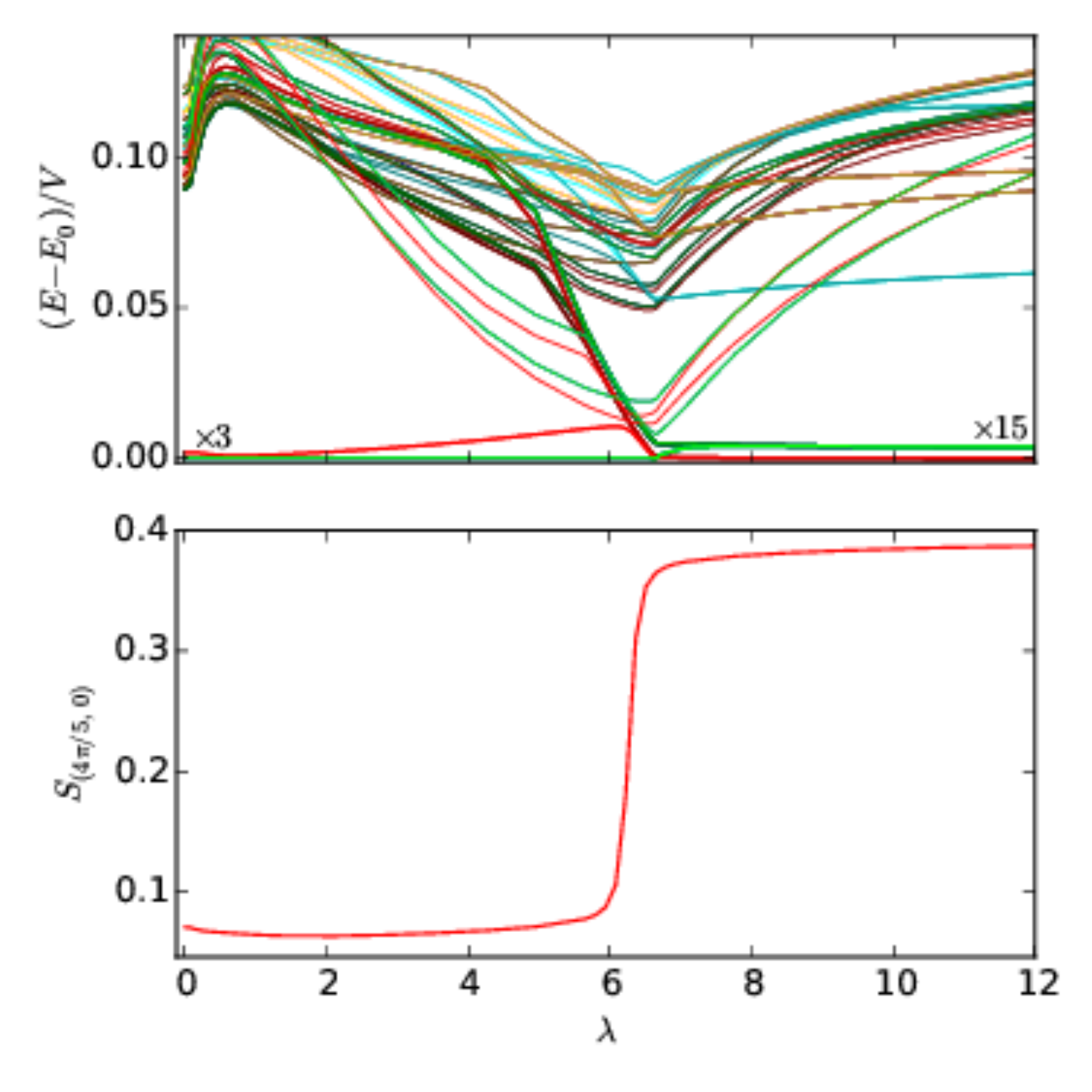}
 \vspace{-1.5in}

  \includegraphics[width=1.2in]{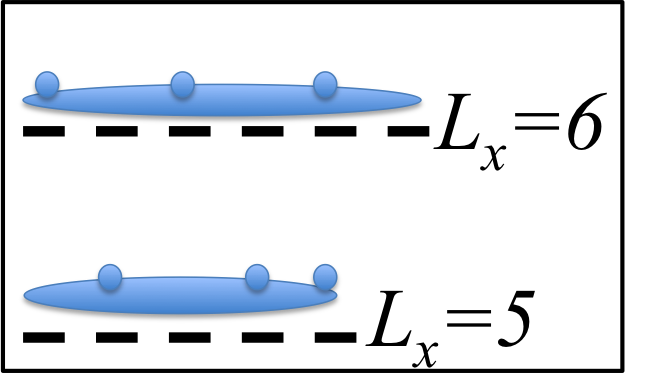}
  \hspace{-2in}
\vspace{0.5in}

  \caption{Top: Energy spectrum of Eq.~(\ref{eq_H}) for $N=10$ plotted as a function of the screening length.  The colors correspond to the same $(K_{x},K_{y})$ as in Fig.~\ref{E_vs_k}.  The transition from an FCI to stripes at $\lambda\approx6.5$ is signaled by a change in degeneracy from 3 to 15.  For $\lambda\gtrsim6.5$ each of the 3 FCI states acquire a 5-fold degeneracy.  Bottom: The ground state structure factor peak plotted for the same parameters as the top panel.  Inset: Schematic of a one-dimensional slice along the $x$-direction (perpendicular to the stripes).  Spheres represent charge excess atop the uniform liquid.  The $L_{x}=5$ ($L_{x}=6$) case fits a  non-integer (integer) number of stripes and shows one of 5 (2) configurations.
}
  \label{E_vs_lambda}
\end{figure}

The top panel of Fig.~\ref{E_vs_lambda} depicts the evolution of the low energy spectrum as a function of the screening length.  The low $\lambda$ limit shows an FCI liquid with 3 degenerate ground states and $\Delta_{\lambda}>0$ up to $\lambda \approx 6.5$ where a transition to a 15-fold degenerate state is found.  We found the same transition in the regime $\lambda = 3.2-6.5$ for $N=6-10$.

The wavevectors and persistence of the underlying 3-fold topological degeneracy of the FCI suggests that the degenerate ground states in Figs.~\ref{E_vs_k}-\ref{E_vs_lambda} for $\lambda > 6.5$ are stripes of quasiparticles.  To show this we first study the spatial symmetry of the charge order.   We then show that the stripe states are best described as ordering of quasiparticles instead of electrons.

\noindent
{\it Stripe Order:}  We use the static structure factor to study the charge order:
\begin{eqnarray*}
  S_{\bf q}   =  \frac{1}{N_c^2} \sum_{j, j'} e^{-i {\bf q} \cdot {\bf r}_{jj'}} \langle \tilde{n}_{j} \tilde{n}_{j'} \rangle,
\end{eqnarray*}
where the tilde indicates projection to the lowest flat band of the single-particle Hamiltonian so that $j$ and $j'$ label unit cells.  We have verified that the low $\lambda$ side of the transition demonstrates uniform ground states, i.e., there are no peaks in $S_{\bf q}$.  But in the large $\lambda$ regime we find evidence for a peak at ${\bf q} =(\pi,0)$ in $S_{\bf q}$ in the thermodynamic limit, i.e., stripes aligned along the $y$-direction.  The unprojected structure factor leads to qualitatively similar results.

The bottom panel of Fig.~\ref{E_vs_lambda} shows the evolution of the structure factor peak with screening length.  For low $\lambda$, $S_{\bf q}$ remains at its background value until an apparently first order transition at $\lambda\approx 6.5$.  (The transition softens for $N<10$) The peak of $S_{\bf q}$ here occurs at, $(4\pi/5,0)$ and $(6\pi/5,0)$.   

\begin{figure}[t]
  \resizebox{1\columnwidth}{!}{\includegraphics{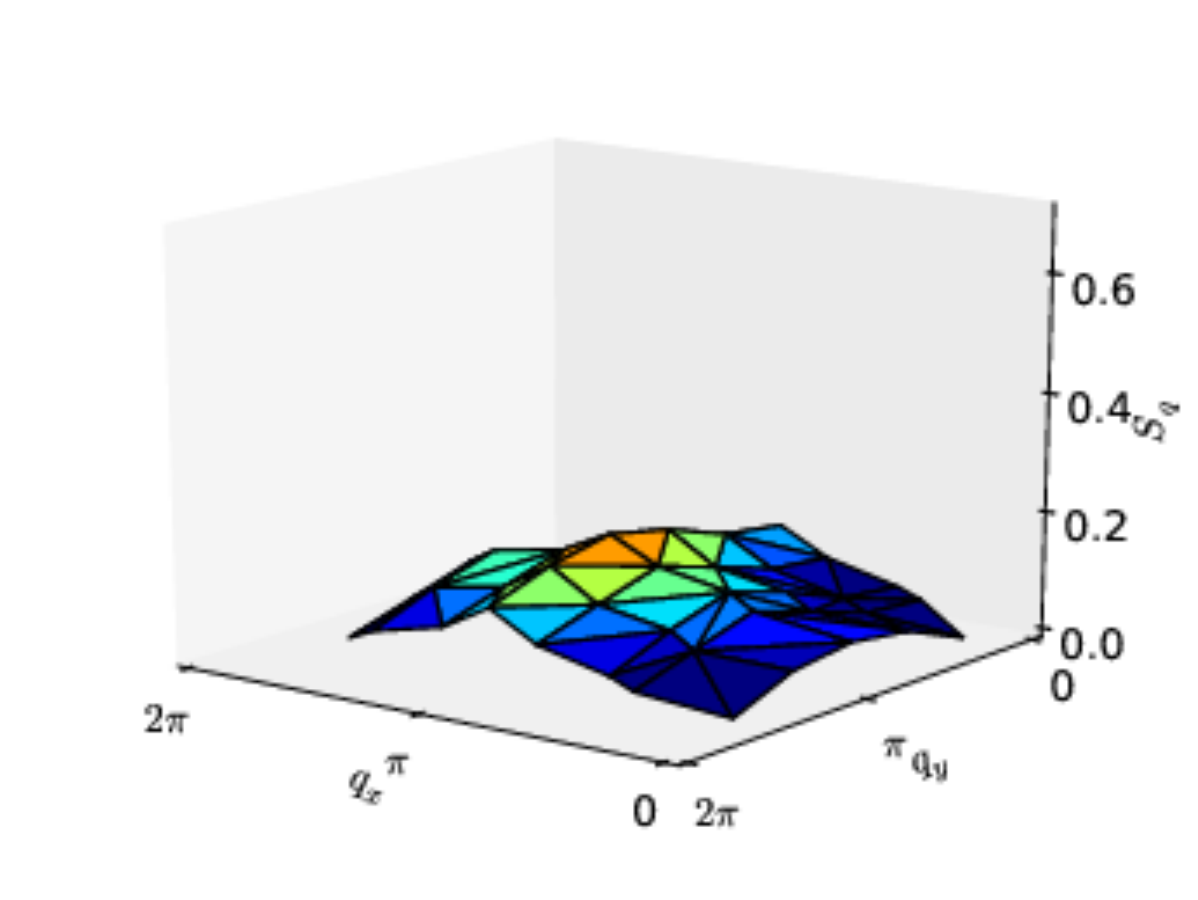}}
  
  \resizebox{1\columnwidth}{!}{\includegraphics{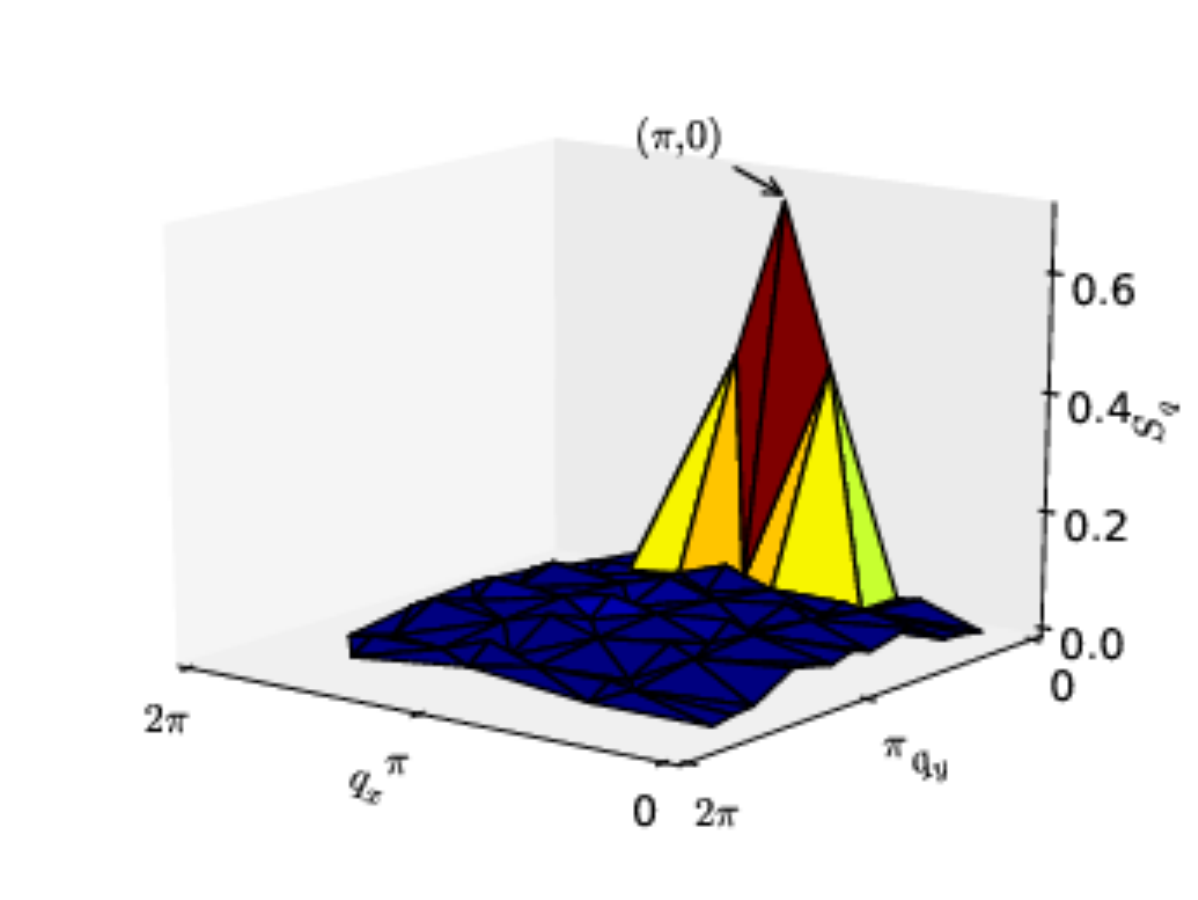}}
  \caption{The top panel plots the ground state structure factor versus wavevector for the uniform FCI for $\lambda =1$ in Eq.~(\ref{eq_H}).  
The bottom panel demonstrates the formation of a stripe state for the Coulomb interaction, $\lambda = \infty$.   }
  \label{sq_3d}
\end{figure}

 Fig.~\ref{sq_3d} plots the structure factor by combining data for $N=8,10$, and $12$.  The top (bottom) panel indicates the FCI (stripe) state.  In the bottom panel, lattices with $N=8$ and $12$ particles show peaks at ${\bf q} =(\pi,0)$, while the $N=10$ data shows peaks at ${\bf q} =(4\pi/5,0)$ and $(6\pi/5,0)$, centered at ${\bf q} =(\pi,0)$.  
 Different system sizes allow different values for the peak.  There is one peak, at $(\pi,0)$, for $L_{x}$ even and two peaks, at $((L_{x}\pm1)\pi/L_{x},0)$, for $L_{x}$ odd.
 We therefore conclude that the peak of $S_{\bf q}$ converges to ${\bf q} =(\pi,0)$ for large system sizes.   
 
$S_{\bf q}$ indicates that the ground states are uniform stripes along the $y$-direction spaced by two lattice constants along the $x$-direction.  The ground state degeneracy can then be explained as the number of ways [$L_{x}$ ($2$) for odd (even) $L_{x}$] to translate the stripes along $x$ (perpendicular to the stripes) and obtain the same state for each of the 3 topological sectors of the FCI.  The even $L_{x}$ case allows stripes to occupy only two distinct sublattices whereas the $L_{x}$ odd case frustrates the two-lattice constant spacing forcing an $L_{x}$ degeneracy (See inset Fig.~\ref{E_vs_lambda}).  Excited states in Fig.~\ref{E_vs_k} then correspond to density modulation along the $y$-direction which have $\Delta_{\lambda}>0$ in our finite-size simulation.

\noindent
{\it Fractionally Charged Quasiparticles:}  We find that the stripe state is a CDW of fractionally charged quasiparticles atop the otherwise uniform FCI liquid.  We note that the charge of the FCI excitations is $e/3$ \cite{bernevig:2011}.  The charge of quasiparticle stripes must therefore be an integer multiple of $e/3$.    To verify this connection we perform flux insertion \cite{neupert:2011, sheng:2011, bernevig:2011} and compute the density in real space to extract the charge of each stripe.  

Laughlin's gauge argument \cite{laughlin:1981} points out that flux inserted along a cylinder axis induces perpendicular current on the surface.  Gauge invariance implies that the periodicity with respect to the flux reveals the charge of quasiparticles because we can make the replacement $eA\rightarrow e^{*}A$, where $A$ is the vector potential and $e^{*}$ is the renormalized charge.  An added condition, $\Delta_{\lambda}>0$, allows for adiabatic spectral flow \cite{tao:1984}.  When both conditions are met they imply a quantized Hall resistance with a value dictated by the charge of the quasiparticles \cite{laughlin:1981,halperin:1982,tao:1984}.  

\begin{figure}[t]
  \resizebox{1\columnwidth}{!}{\includegraphics{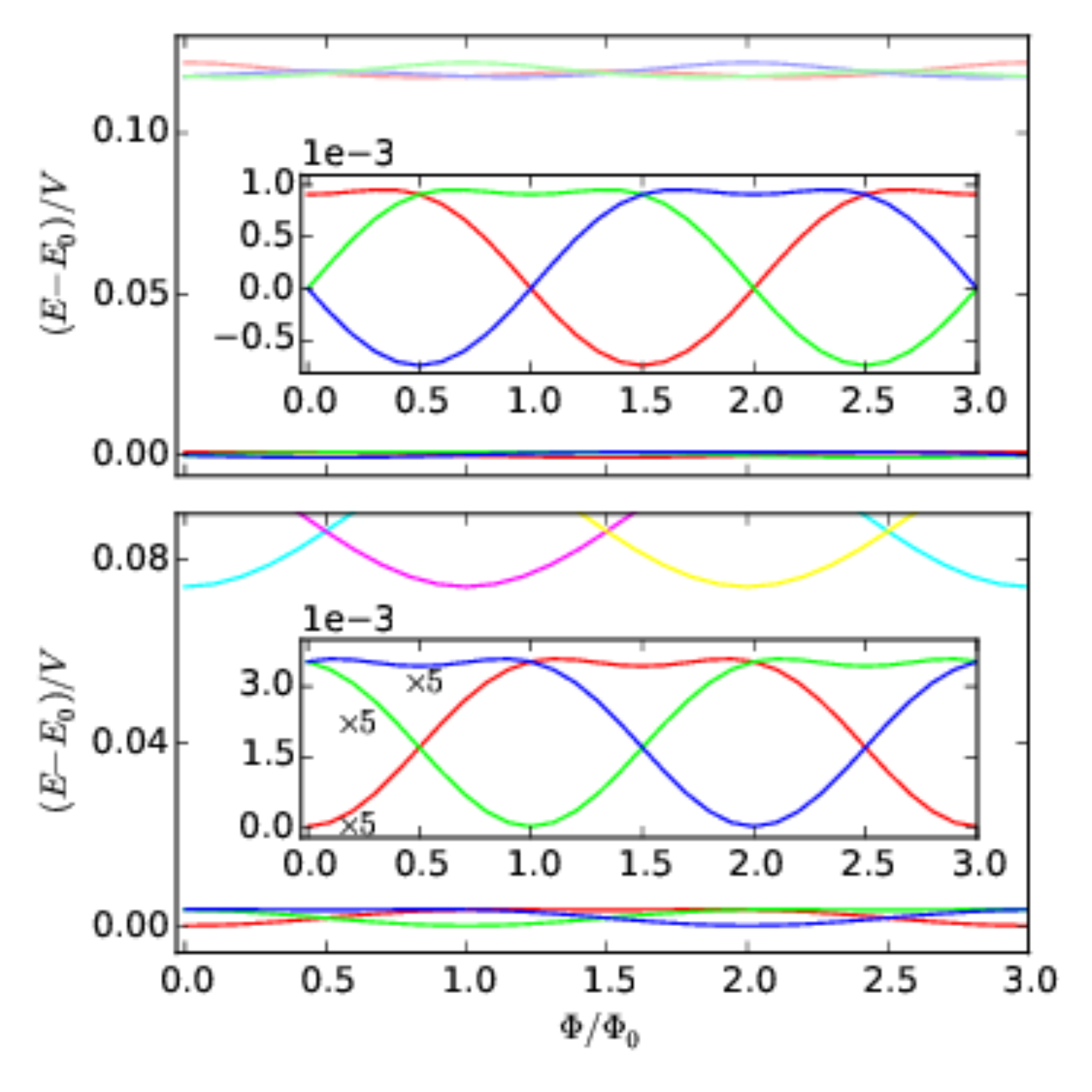}}
  \caption{Spectral flow of the FCI ($\lambda =1$, top panel) and stripe state ($\lambda=\infty$, bottom panel) as a function of the flux in units of $\Phi_{0}=h/e$ for $N=10$.  Both panels show a $\Phi_{0}$ spectral periodicity and a $3\Phi_{0}$ eigenstate periodicity with respect to the flux.   The spectral flows are adiabatic within the ground state manifold.  The insets zoom in on the ground state manifolds.  The colors correspond to the same $(K_{x},K_{y})$ as in Fig.~\ref{E_vs_k}.   }
  \label{fig_5}
\end{figure}

The top panel of Fig.~\ref{fig_5} shows adiabatic spectral flow with respect to flux insertion in the FCI regime but with $\lambda=1$.  Here a test flux is inserted along the $x$ direction.  The spectra (eigenstates) show $\Phi_{0}$ ($3\Phi_{0}$) periodicity in $\Phi$.  $\Delta_{\lambda}>0$ and the $3\Phi_{0}$ eigenstate periodicity imply a Hall conductivity quantized at $1/3$.  Our results therefore show that the FCI is favored for interactions with a non-zero screening length, $\lambda \lesssim 6.5$. 

The bottom panel of Fig.~\ref{fig_5} shows representative flux insertion data in the Coulomb limit.  Here the flux pumps charge parallel to the stripes.  The eigenstates of the stripe phase show a $3\Phi_{0}$ periodicity.  Gauge invariance therefore implies that flux evolution of the ground state is consistent with stripes of quasiparticles with charge $e/3$.  There also appears to be a spectral gap and adiabatic spectral flow consistent with fractionally quantized Hall conductivity in the stripe phase.  Our finite-size simulations cannot rule out the possibility that the gap collapses in the thermodynamic limit due to long wavelength fluctuations in the stripe phase.

To conclusively show that the stripes are fractionally charged quasiparticles atop an otherwise uniform state, we compute the charge of a single stripe.  We add a weak symmetry breaking term to Eq.~(\ref{eq_H}):  $\sum_{j} \epsilon_j  \tilde{n}_j $.  For $\epsilon_j\ll \Delta_{\lambda}$ we obtained the same stripe density regardless of our choice for $\epsilon_{j}$, i.e., the stripes spontaneously break the $C_{4}$ lattice symmetry.  We also find that 3 of the degenerate stripe states at $K_{FCI}$ are uniform in the single-particle density, i.e., $F_{\rho_{\bf Q}}\sim 1+\mathcal{O}(\rho_{{\bf Q}}^{2})$ in the wavefunction.  This is in contrast to the other stripe states in the degenerate manifold showing charge modulation, $F_{\rho_{\bf Q}}\sim \rho_{{\bf Q}}+\mathcal{O}(\rho_{{\bf Q}}^{2})$.  

The inhomogeneous density allows us to compute the charge of one stripe:
$
N_{qp}e^{*}/e=\sum_{j\in R} \langle \tilde{n}_{j}-\rho_{0} \rangle,
$
where $N_{qp}$ is the number of quasiparticles, $\rho_{0}=1/3$ is the density of the uniform liquid, and the region $R$ defines summation over a single stripe by choosing $j$'s such that $\langle \tilde{n}_{j} \rangle >\rho_{0}$.  For the uniform stripes studied here this condition implies that we sum over the entire stripe.  We find $N_{qp} e^{*}/e$ to be a multiple of 1/3 within numerical accuracy for all system sizes studied by computing the charge within the subspace of stripe states showing charge modulation.  For example, for $N=8$ we find $N_{qp}e^{*}/e=4/3$.  Since the FCI quasiparticles have charge 1/3 it is natural to conclude that we have $N_{p}=4$ and $e^{*}/e=1/3$.  We have tested that this results is robust against different choices for $\epsilon_j$.  We have therefore found that increasing $\lambda$ orders quasiparticles atop the otherwise uniform liquid into stripes.  

\noindent
{\it Constraints on Effective Theories:}  Constraints on effective theories illustrate the difference between CDWs of electrons and CDWs of quasiparticles.  A minimal effective model must reproduce the lowest energies and momenta.  Truncation of Eq.~(\ref{eq_H}) does not accomplish this.  For example, a classical model derived by dropping off-diagonal terms in Eq.~(\ref{eq_H}), $H_{\text{c}}$, leads to a non-degenerate CDW in the ground state, e.g., at $(K_{x},K_{y})=(2,4)$ for N=8.  Adding the largest off-diagonal terms: $ \tilde{c}_{x,y}^{\dagger} \tilde{n}_{x,y\pm1} \tilde{c}_{x\pm1,y}+h.c.$, to $H_{\text{c}}$, leads to a CDW with degeneracies and momentum sectors that are very different from the full model.  

An effective theory of quasiparticles must also satisfy an exact constraint of fractional charge.  Consider, for example, an effective model built from  anisotropic hopping and interaction terms: $-t \sum_{\langle j,j' \rangle} \vert \hat{r}_{j,j'}\cdot \hat{y} \vert( \tilde{c}_j^{\dagger} \tilde{c}_{j'}^{\vphantom{\dagger}} +h.c. ) + \sum_{j\neq j'} (1-c \vert \hat{r}_{j,j'}\cdot \hat{y}\vert)\tilde{n}_{j} \tilde{n}_{j}/(2r_{jj'})$, where $t$ and $c<1$ are fitting parameters and $\hat{r}_{j,j'}$ and $\hat{y}$ are unit vectors.  For large $c$ this theory can be thought of as weakly coupled Luttinger liquids \cite{mukhopadhyay:2001}.   We have checked that the low energy states of this model do not order at the same wavevectors as Eq.~(\ref{eq_H}).  But more importantly, the stripes defined this way can be fractionalized at \emph{any} value.  The effective theory must also define an array of FCI quasiparticles with a charge that is a rational fraction \cite{kane:2002,klinovaja:2013,neupert:2014,klinovaja:2014a}.

\noindent
{\it Summary:} We tested a model of FCIs that includes \emph{a priori} screening of the underlying Coulomb interaction.  Adiabatic flux insertion shows that the FCI survives for sizable screening lengths.  The perseverance of the FCI state (and its quasiparticles) allowed a transition to an intriguing stripe phase of fractionally charged quasiparticles.  Our results show that some of the rich structure considered in the FQH regime, e.g., stripes of quasiparticles, manifests in the FCI regime.    
 
\begin{acknowledgements}
We acknowledge support from ARO (W911NF-12-1-0335), AFOSR (FA9550-11-1-0313), and DARPA-YFA.
\end{acknowledgements}

\bibliographystyle{apsrev4-1}
\bibliography{references.bib}

\end{document}